\documentstyle[12pt]{article}
\title{  {\normalsize{{\hskip 11cm}  BIHEP-TH-95-28}}\\
     Electroweak Penguin Contributions in $B^-(\overline {B_d^0})
   \rightarrow  PP$ and $PV $ Decays Beyond Leading Logarithms   }
\author{
	 Dongsheng Du$\rm ^{a,b}$ and  Libo Guo$\rm ^{b,c}$
	\thanks{E-mail address: Duds@bepc3.ihep.ac.cn;
	 $~~$Guolb@bepc3.ihep.ac.cn}\\
	 $\rm ^a$ \it{CCAST(World Laboratory),
	 P.O.Box 8730, Beijing 100080, China}\\
	 $\rm ^b$ \it{Institute of High Energy Physics, Chinese Academy
	 of Sciences}\\
	 \it{P.O.Box  918(4),  Beijing, 100039, P. R. China}
	\thanks{mailing address.}\\
	$\rm ^c$ \it{Department of Physics, Henan Normal University}\\
	  \it{Xinxiang, 453002, Henan, P.R. China} }
\date{}
\begin{document}

\maketitle
13.25.Hw,  13.40Hq
\begin{abstract}

Using the next-to-leading order low energy effective Hamiltonian for
$|\Delta B|=1$, $\Delta C=\Delta U=0$ transitions, the contributions of
electroweak penguin operators in $ B^-(\overline {B_d^0})\rightarrow  PP$
and $PV$ decays are estimated in the standard model. We find that, for some
channels, the electroweak penguin effects can enhance or reduce the QCD
penguin and/or tree level contributions by at least $30\%$, and can even
play dominant role.
\end{abstract}

\newpage
\section{Introduction}

In a recent publication [1], the QCD penguin contributions to decay rates and
direct CP violating rate asymmetries are systematically investigated using the
low energy effective Harmiltonian [2] including next-to-leading order QCD
correction for $|\Delta B|=1$, $\Delta C=\Delta U=0$ transitions. Since this
Hamiltonian has been generalized by Bura et al through the inclusion of
electroweak penguin operators [3,4,5], now we are in a position to analyze not
only the contributions of the QCD penguin operators beyond the leading
logarithmic approximation, but also the contributions of the electroweak
penguins. Naively, people believe that the electroweak penguin contributions
are suppressed by a factor of $\alpha_{em}/\alpha_s \sim {\cal O}(10^{-2})$
relative to QCD corrections [7,8], so the electroweak penguin corrections
comparing with the QCD penguin may be negligible. However, this is not always
true.

As is well-known from the anatomy of the penguin-dominated quautity
$\epsilon^\prime /\epsilon$ describing direct CP violation in the K-meson
system, the electroweak penguin contributions can become important and can
even compete with QCD penguin operator contributions in the presence of a
heavy top-quark [5,6]. But how about in $ B^-(\overline {B_d^0}) \rightarrow
PP $ and $PV$ decays?

In this paper, by applying the BSW model [9] based on the factorization
assumption, we calculate the branching ratios and rate asymmetries in
$B^-(\overline {B_d^0}) \rightarrow  PP$ and $PV$ decays, and find that,
for some decay channels, the electroweak penguin contributions can be as
large as $30\% \sim 95\%$ in decay width and the CP-asymmetries can be
larger than $30\%$ in $CP$ asymmetries.

The paper is organized as follows: In section 2, we give a brief description
of the low energy effective Hamiltonian including
both leading and next-to-leading order QCD corrections and leading
order corrections in the QED coupling constant and the explicit cancellation
of the renormalization scheme dependence arising beyond the leading
logarithmic approximation. In section 3, we then apply the effective
Hamiltonian discussed in section 2 and the BSW model in $B^-(\overline {B_d^0})
\rightarrow  PP$ and $PV$  decays to calculate the weak decay amplitudes.
The numerical results for the branching ratios and CP-asymmetries are
disscussed in section 4. At the end, we give a brief summary.

\section{ The Effective Hamiltonian Beyond Leading Logarithms and the
Cancellation of Renormalization Scheme Dependence}

As discussed in ref. [5], the next-to-leading order low energy effective
Hamiltonian describing $\Delta B=-1$, $\Delta C=\Delta U=0$ transitions takes
the following form at the renormalization scale $\mu=m_b$
\begin{eqnarray}
{\cal }H_{eff}(\Delta B=-1) = \frac{G_F}{\sqrt{2}}\left[\sum_{q=u,c}v_q
			      \left\{ Q_1^q C_1(\mu)+Q_2^qC_2(\mu)
			      +\sum_{k=3}^{10} Q_k C_k(\mu)\right\}\right]
\end{eqnarray}
where $C_k(\mu)$ (k=1,$\cdots$,10) are the Wilson coefficients which  are
calculated in renormalization group improved pertubation theory and include
leading and next-to-leading order QCD corrections and leading order QED
corrections. The CKM factors $v_q$ are defined as
$$
v_q=\left\{
	  \begin{array}{ll}
	  V_{qd}^\ast V_{qb} & \mbox{for $b \rightarrow d$ transitions}\\
	  V_{qs}^\ast V_{qb} & \mbox{for $b \rightarrow  s$ transitions.}
	  \end{array}
	  \right.
\eqno(2)$$
Using the Wolfenstein parametrization [10] in which the CKM matrix is
parameterized by $A$, $\lambda$, $\eta$ and $\rho$, we have
$$
v_u=\left\{
	  \begin{array}{ll}
	  A\lambda^3(\rho-i\eta) & \mbox{for $b \rightarrow d$ transitions}\\
	  A\lambda^4(\rho-i\eta) & \mbox{for $b \rightarrow  s$ transitions.}
	  \end{array}
	  \right.
\eqno(3)$$
$$
v_c=\left\{
	  \begin{array}{ll}
	 -A\lambda^3 & \mbox{for $b \rightarrow d$ transitions}\\
	 A\lambda^2(1+i\eta\lambda^2) &\mbox{for $b\rightarrow s$ transitions.}
	  \end{array}
	  \right.
\eqno(4)$$
In our numerical calculation, we use $\lambda=0.22$, $A=0.8$, $\eta=0.34$,
$\rho=-0.12$ which are obtained from the fit to the experimental data [11].
The operators $Q_1^u$, $Q_2^u$, $Q_3$, $\cdots$, $Q_{10}$ are given as
follows:   $$
\begin{array}{ll}
Q_1^u=(\bar{q}_{\alpha}u_{\beta})_{V-A}(\bar{u}_{\beta}b_{\alpha})_{V-A} &
Q_2^u=(\bar{q}u)_{V-A}(\bar{u}b)_{V-A}\\
Q_3=(\bar{q}b)_{V-A}\sum_{q'}(\bar{q'}q')_{V-A} &
Q_4=(\bar{q}_{\alpha}b_{\beta})_{V-A}\sum_{q'}
    (\bar{q'}_{\beta}q'_{\alpha})_{V-A}\\
Q_5=(\bar{q}b)_{V-A}\sum_{q'}(\bar{q'}q')_{V+A} &
Q_6=(\bar{q}_{\alpha}b_{\beta})_{V-A}\sum_{q'}
    (\bar{q'}_{\beta}q'_{\alpha})_{V+A}\\
Q_7=(\bar{q}b)_{V-A}\sum_{q'}\frac{3}{2}e_{q'}(\bar{q'}q')_{V+A} &
Q_8=(\bar{q}_{\alpha}b_{\beta})_{V-A}\sum_{q'}\frac{3}{2}e_{q'}
    (\bar{q'}_{\beta}q'_{\alpha})_{V+A}\\
Q_9=(\bar{q}b)_{V-A}\sum_{q'}\frac{3}{2}e_{q'}(\bar{q'}q')_{V-A} &
Q_{10}=(\bar{q}_{\alpha}b_{\beta})_{V-A}\sum_{q'}\frac{3}{2}e_{q'}
     (\bar{q'}_{\beta}q'_{\alpha})_{V-A}\\
\end{array}
\eqno(5)$$
Where $Q_1^q$ and $Q_2^q$ (q=u,c) are current-current operators, for q=c case,
the two operators $Q_1^c$ and $Q_2^c$ are obtained through making
substitution  $u \rightarrow c$  in $Q_1^u$ and $Q_2^u$ respectively;
the operators $Q_3$,$\cdots$,$Q_6$ represent the so-called QCD penguin
operators, whereas $Q_7,\cdots\,Q_{10}$ denote the electroweak penguin
operators which will be of special interest to us in the present paper.
$q'$ is running over the quark flavours being active at the $\mu=
{\cal O}(m_b)$ scale  $(q'\in \left\{ u, d, c, s, b \right\})$ ,
$e_{q'}$ are the corresponding quark charges and q=d or s for
$ b\rightarrow  d$ or s transitions respectively.
The indices $\alpha$, $\beta$ are  $SU(3)_c$ colour indices .

Beyond the leading logarithmic approximation, the Wilson coefficients
$C_k(\mu )$ ( k = 1, $\cdots$, 10 ) obtained by solving the renormalization
group equation depend both on the form of the operator basis (5) and on the
renormalization scheme used. Consequently, the tree level penguin transition
matrix elements or amplitudes calculated by $H_{eff}$ are also scheme
dependent. However, the physical quantities, of course, should be
renormalization scheme independent if one handels the hadronic matrix
elements correctly. In order to cancel these scheme dependence, we introduce
the renormalization scheme independent Wilson coefficient function [2,5,12]:
$$
{\bf \bar{C}}(\mu)=(\hat{1}+\frac{\alpha_s(\mu)}{4\pi}\hat{\gamma}_s^T
		  +\frac{\alpha_{em}(\mu)}{4\pi}\hat{\gamma}_e^T)
		   \cdot {\bf C}(\mu)
\eqno(6)
$$
where $\hat{\gamma}_{s,e}$ are obtained from one-loop matching conditions,
and treat the matrix elements to one-loop level [12,13,21]. These one-loop
matrix elements can be rewritten in terms of the tree-level matrix
elements $<Q_j>_0$ as
$$
<{\bf Q}^T(\mu)>=<{\bf Q }^T>_0 \cdot \left[\hat{1}
	+ \frac{\alpha_s(\mu)}{4\pi}\hat{m}^T_s(\mu)
	+ \frac{\alpha_{em}}{4\pi}\hat{m}^T_e(\mu)\right],
\eqno(7)
$$
which define matrices $\hat{m}_s(\mu)$ and $\hat{m}_e(\mu)$.
In eq.(6) and (7), ${\bf C}(\mu)$, ${\bf \bar{C}}(\mu)$ and
${\bf Q}(\mu)$ are all column matrix. Combining (1) with (6) and
(7), we obtain the following renormalization scheme independent
transition amplitude:
$$
  \begin{array}{rl}
  <H_{eff}(\Delta B=-1)>&=\displaystyle\frac{G_F}{\sqrt{2}}\left[
  \displaystyle\sum_{q=u,c}v_q\left\{<Q_1^q>_0 C_1^{eff}(\mu)+
			<Q_2^q>_0 C_2^{eff}(\mu)\right.\right.\\[4mm]
		       &~+\left.\left.\displaystyle\sum_{k=3}^{10}
			<Q_k>_0 C_k^{eff}(\mu)\right\}\right],
		      \end{array}
   \eqno(8)
$$
where $C_k^{eff}$ are defined as
$$
\begin{array}{rl}
C_1^{eff}&=\bar{C}_1,~~~~~~~~~~~~~~~~~~~~~C_2^{eff}=\bar{C}_2 \\
C_3^{eff}&=\bar{C}_3(\mu)-P_s/3,~~~~~~~C_4^{eff}=\bar{C}_4(\mu)+P_s \\
C_5^{eff}&=\bar{C}_5(\mu)-P_s/3,~~~~~~C_6^{eff}=\bar{C}_6(\mu)+P_s \\
C_7^{eff}&=\bar{C}_7(\mu)+P_e,~~~~~~~~~C_8^{eff}=\bar{C}_8(\mu) \\
C_9^{eff}&=\bar{C}_9(\mu)+P_e,~~~~~~~~~C_{10}^{eff}=\bar{C}_{10}(\mu) \\
\end{array}
\eqno(9)
$$
where $P_{s,e}$ are given by
$$
  \begin{array}{rl}
   P_s&=\displaystyle\frac{\alpha_s}{8\pi}\left[\frac{10}{9}-
     G(m_q,q^2,\mu)\right]\bar{C}_2(\mu),\\[4mm]
   P_e&=\displaystyle\frac{\alpha_{em}}{9\pi}\left[\frac{10}{9}-
	    G(m_q,q^2,\mu)\right]\left(3\bar{C}_1(\mu)
	    +\bar{C}_2(\mu)\right),\\[4mm]
G(m,q^2,\mu)&=-4\displaystyle\int_0^1 x(1-x) ln\displaystyle
	      (\frac{m^2-x(1-x)q^2}{\mu^2}) dx
\end{array}
\eqno(10)
$$
here q denotes the momentum of the virtual gluons and photons or $Z^0$ bosons
appearing in the QCD and QED time-like matrix elements. For the details of
this calculation, the reader is referred to refs.[12,21].
In the numerical calculation, we will use $q^2=m_b^2/2$
which represents the average value.

If we take $m_t=174GeV$, $m_b=5.0GeV$, $\alpha_s(M_z)=0.118$,
$\alpha_{em}(M_z)=1/128$ , we have the numerical values of
the renormalization scheme independent Wilson coefficients
$\bar{C}_i$ at $\mu=m_b$ scale as follows [14]:
$$
\begin{array}{rl}
\bar{C}_1&=-0.3125,~~~~\bar{C}_2=1.1502,~~~~\bar{C}_3=0.0174,\\
\bar{C}_4&=-0.0373,~~~~\bar{C}_5=0.0104,~~~~\bar{C}_6=-0.0459,\\
\bar{C}_7&=-1.050\times 10^{-5},~~~~~~~~~\bar{C}_8=3.839\times10^{-4},\\
\bar{C}_9&=-0.0101,~~~~~~~~~~~~~~~~~\bar{C}_{10}=1.959\times 10^{-3},\\
\end{array}
\eqno(11)
$$

\section{ Decay Amplitudes in BSW Model}

To work out the decay amplitudes, we follow Bauer, Stech and Wirbel [9].
With the help of the factorization hypothesis [15], the three-hadron matrix
elements $<XY|H_{eff}|B>$, that is the decay amplitude,
can be factorized into a sum of products of two current
matrix elements $<X|J_1^{\mu}|0>$ and $<Y|J_{2\mu}|B>$.
The former matrix
elements are simply given by the corresponding decay constants
$f_X$ and $g_X$ [16]:\\
$$
\begin{array}{rl}
<0|J_{\mu}|X(0^-)>&=i f_X k_{\mu}\\
<0|J_{\mu}|X(1^-)>&=M_X g_X \epsilon_{\mu}
\end{array}
\eqno(12)
$$    \\
\noindent for a pseudoscalar and a vector meson with the polarization vector
$\epsilon_{\mu}$ respectively, where
$J_{\mu}=V_{\mu}-A_{\mu}$ is
the usual colour-singlet V-A current. The latter kind of matrix elements
can be expressed in terms of Lorentz-scalar form factors [9,16]:

$$
\begin{array}{rl}
<X(0^-)|J_{\mu}|B>&=\displaystyle \left[\left(k_B+k_X \right)_{\mu}-
	   \frac{M_B^2-M_X^2}{k^2} k_{\mu} \right] F_1^{BX}(k^2)\\[4mm]
	      & +\displaystyle\frac{M_B^2-M_X^2}{k^2} k_{\mu} F_0^{BX}(k^2)\\[4mm]
<X(1^-)|J_{\mu}|B>&=\displaystyle \frac{2}{M_B+M_X} \epsilon_{\mu \nu
		     \rho \sigma}\epsilon^{*\nu} k_B^{\rho} k_X^{\sigma}
		      V^{BX}(k^2) \\[4mm]
		& +i \displaystyle\left(M_B+ M_X \right)\left[ \epsilon_{\mu}^*-
		  \frac{\epsilon^*\cdot {k}}{k^2} k_{\mu}\right]A_1(k^2)\\[4mm]
	   & -i \displaystyle\frac{\epsilon^* \cdot k}{M_B+M_X}[(k_B+k_X)_{\mu}-\frac
		{M_B^2-M_X^2}{k^2} k_{\mu}] A_2(k^2) \\[5mm]
	   & +i \displaystyle\epsilon^* \cdot k \frac{2 M_X}{k^2} k_{\mu} A_0(k^2)
\end{array}
\eqno(13)
$$  \\
\vspace{0.3cm}
\noindent where $ k^{\mu}=k_B^{\mu}-k_X^{\mu}$ and $M_B$, $M_X$, $M_Y$ are the
masses of meson B, X and Y, respectively.
The form factors $F_{0,1}(k^2), V(k^2), A_{0,1,2}(k^2)$ are related to
the form factors $F_{0,1}(0), V(0), A_{0,1,2}(0)$ defined in ref. [9] by

$$
\begin {array}{rl}
F_0(k^2)~&=~\displaystyle\frac{F_0(0)}{1-k^2/m^2(0^+)}~~~~~~~
F_1(k^2)~=~\displaystyle\frac{F_1(0)}{1-k^2/m^2(1^-)}\\ [4mm]
A_0(k^2)~&=~\displaystyle\frac{A_0(0)}{1-k^2/m^2(0^-)}~~~~~~~
 A_1(k^2)~=~\displaystyle\frac{A_1(0)}{1-k^2/m^2(1^+)}\\[4mm]
A_2(k^2)~&=~\displaystyle\frac{A_2(0)}{1-k^2/m^2(1^+)}~~~~~~~
 V(k^2)~=~\displaystyle\frac{V(0)}{1-k^2/m^2(1^-)}       \\
\end{array}
\eqno(14)
$$

\noindent On the other hand, there are additional contributions from (V+A)
penguin
operators, for example $Q_5$, $Q_6$, $Q_7$ and $Q_8$ in (5).
Using the equation of motion, this kind of
matrix elements can be rewritten in terms of those involving usual (V-A)
currents [17]. After a straightforward calculation, we can obtain
three different kinds of expressions relevant to
$M^{XY}\equiv~<X|J_{1\mu}|0><Y|J_2^{\mu}|B>$:\\

\noindent {\it case 1}~~   $ J_X^P=0^-$ and $J_Y^P=0^-$:
$$
\begin{array}{rl}
M^{XY}=&-i (M_B^2-M_Y^2)f_XF_0^{BY}(M_X^2)
\end{array}
\eqno(15)
$$
{\it case 2}  ~~ $ J_X^P=0^-$ and $J_Y^P=1^-$:
$$\begin{array}{rl}
M^{XY}=&2 M_Y f_X A_0^{BY}(M_X^2)(\epsilon_Y^*\cdot k_X)
\end{array}
\eqno(16)  $$
{\it case 3} ~~  $ J_X^P=1^-$ and $J_Y^P=0^-$:
$$\begin{array}{rl}
M^{XY}=&2 M_X g_X F_1^{BY}(M_X^2)(\epsilon_X^*\cdot k_B)
\end{array}
\eqno(17) $$
Conveniently, we define
$$\begin{array}{rl}
a_{2i-1}&\equiv\displaystyle{C}_{2i-1}^{eff}
	 +\frac{{C}_{2i}^{eff}}{3},
{}~~~~~a_{2i}\equiv\displaystyle{C}_{2i}^{eff}
	 +\frac{{C}_{2i-1}^{eff}}{3},
{}~~~~(i=1,\cdots,5)
\end{array}
\eqno(18)$$
Using these formulas, we can give the decay amplitudes of
$B^-(\overline {B_d^0}) \rightarrow PP$ and $PV$ decays. As an example, we
give the result of $<\pi^-\pi^0|H_{eff}|B_u^->$ in the following:
$$
\begin{array}{rl}
<\pi^-\pi^0|H_{eff}|B_u^->~=&\displaystyle\frac{G_F}{\sqrt{2}}\displaystyle\left[v_u(a_1
       M^{\pi^-\pi^0}_{uud}+a_2 M^{\pi^-\pi^0}_{duu})\right.\\[4mm]
 +  &\displaystyle\sum_{q=u,c} v_q\left\{(-a_4-\frac{3}{2} a_7
 +\frac{3}{2} a_9+\frac{a_{10}}{2}\right.\\[4mm]
   -&\displaystyle\frac{M_{\pi^0}^2}{m_d(mb-m_d)} (a_6-\frac{a_8}{2}))
M^{\pi^-\pi^0}_{uud}\\[4mm]
 +  &\displaystyle\left.\left.(a_4+a_{10}+\frac{2
M_{\pi^-}^2}{(m_b-m_u)(m_u+m_d)}(a_6+a_8)) M^{\pi^-\pi^0}_{duu}\right\}\right]
\\
\end{array}
\eqno(19)
$$
where $M^{\pi^-\pi^0}_{uud}$ and $M^{\pi^-\pi^0}_{duu}$ are defined as follows:
$$
\begin{array}{rl}
M^{\pi^-\pi^0}_{uud}\equiv~&<\pi^0|(\bar{u}u)_{V-A}|0><\pi^-|(\bar{d}b)_{V-A}|B_u^->\\[4mm]
	=~&-i(M_B^2-M_{\pi^-}^2)f_{\pi^0}F^{B\pi}_0(M_{\pi^0}^2)\\[4mm]
M^{\pi^-\pi^0}_{duu}\equiv~&<\pi^-|(\bar{d}u)_{V-A}|0><\pi^0|(\bar{u}b)_{V-A}|B_u^->\\[4mm]
	=~&-i(M_B^2-M_{\pi^0}^2)f_{\pi}F^{B\pi}_0(M_{\pi^-}^2)/\sqrt{2}\\[4mm]
\end{array}
\eqno(20)
$$
where $f_{\pi}$ and $f_{\pi^0}$ are the decay constants of the $\pi$ and
$\pi^0$ meson, respectively,
$F_0$ is the form factor(see Appendix).

\section{ NUMERICAL RESULTS AND DISCUSSIONS}

In the $B$ rest frame, the two body decay width is [18]
$$
\begin{array}{ll}
\Gamma(B\rightarrow XY)~=&\displaystyle\frac{1}{8\pi}|<XY|H_{eff}|B>|^2
	  \frac{|p|}{M_B^2}
\end{array}
\eqno(21)
$$
where
$$
\begin{array}{rr}
|p|=&\displaystyle\frac{[(M_B^2-(M_X+M_Y)^2)(M_B^2-(M_X-M_Y)^2)]^{1/2}}{2M_B}
\end{array}
\eqno(22)
 $$
is the magnitude of the momentum of the particle X or Y. The corresponding
branching ratio is given by
$$
\begin{array}{ll}
BR(B\rightarrow XY)=&\displaystyle\frac{\Gamma(B\rightarrow XY)}
{\Gamma^B_{tot}}
\end{array}
\eqno(23)
$$
In our numerical calculation, we take $\Gamma^{B_u^-}_{tot}=4.273\times
10^{-13}$GeV and  $\Gamma^{B_d^0}_{tot}=4.387\times 10^{-13}$ GeV[19].

The CP-asymmetry ${\cal A}_{CP}$ for the charged $B$ system is defined as
$$
\begin{array}{ll}
{\cal A}_{CP}=&\displaystyle\frac{\Gamma (B^-\rightarrow f)-\Gamma
(B^+\rightarrow \bar{f})}
  {\Gamma (B^-\rightarrow f)+\Gamma (B^+\rightarrow \bar{f})}
\end{array}
\eqno(24)
$$
While for the neutral B-meson system,
$$
\begin{array}{ll}
{\cal A}_{CP}=&\displaystyle\frac{\Gamma (\overline{B_0}\rightarrow f)-
\Gamma (B_0\rightarrow \bar{f})}
  {\Gamma (\overline{B_0}\rightarrow f)+\Gamma (B_0\rightarrow \bar{f})}
\end{array}
\eqno(25)
$$
if f is CP-eigenstate, eq(25) can be simplified as
$$
\begin{array}{ll}
{\cal A}_{CP}=&\displaystyle\frac{1-|\xi|^2-2 x Im \xi }
  {(1+|\xi|^2)(1+x^2)}
\end{array}
\eqno(26)
$$
where $x$ and $\xi$ are defined in [22].
In our numerical calculation, we take $x_d=0.71$.
The numerical results of electroweak penguin contributions to the branching
ratios and CP asymmetries are collected in table 1-8, where ``$QCD+QED$''
means the branching ratios and CP asymmetries with full QCD and QED penguin
contributions, ``QCD'' with only QCD penguin contributions,
$\delta \Gamma/\Gamma$ represents the enhancement percentage of the
electroweak penguin contributions to the decay width, $\delta {\cal A}/{\cal
A}$
represents the enhancement percentage of the electroweak penguin contributions
to rate asymmetry.
All the parameters needed in the calculations, such as the meson decay
constants, form factors and pole masses etc, are listed in the appendix.

{}From table 1-8, we can see the following features:

{}~~~1)  For $B^-\rightarrow\phi\pi^-$ and $ \bar{B}_d^0\rightarrow\phi\phi^0$,
 $\eta$, $\eta'$, the electroweak penguin contributions to the branching
ratios are dramatically large, they can enhance the branching ratio by 3
orders in magnitude. This is because the relation $C_3(mb)+C_5(mb)\approx
-1/3(C_4(mb)+C_6(mb))$ which could lead to a strong cancellation between
the $QCD$ penguin matrix elements in the decay amplitudes.

 For $B^-\rightarrow K^-\omega$, $K^-\rho^0$, $K^{-*}\eta$ and $\bar{B}_d^0
\rightarrow \bar{K^0}\eta$, $\bar{K^0}\rho^0$, $\bar{K^0}\omega$, the
electroweak penguin effects compete with or dominate over $QCD$ penguin
and/or tree level contributions in branching ratios. For example, the
electroweak penguin effects can enhance the branching ratio by a factor
of $3$ in magnitude for $B^-\rightarrow K^-\rho^0$. This is becausse the
particular structure of
$\rho^0=\frac{1}{\sqrt{2}}(\bar{u}u-\bar{d}d)$ which could lead to a strong
cancellation between the $\bar{u}u$ and $\bar{d}d$ contributions in $QCD$
penguin matrix elements.

 For $B^-\rightarrow K^-\pi^0$, $K^-\eta$ ,$K^{-*}\eta'$, $K^{-*}\pi^0$
and $\bar{B}_d^0\rightarrow \pi^0\pi^0$, $\pi^0\rho^0$, $\bar{K^0}^*\pi^0$,
the electroweak penguin effects are non-negligible in branching ratiios,
they can enhance or reduce the $QCD$ and/or tree level contributions by a
percentage larger than $30\%$.

2) For $B^-\rightarrow \phi\pi^-$, $\pi^-J/\psi$ and $\bar{B}_d^0\rightarrow
\bar{K^0}J/\psi$, the electroweak penguin contributions are the only source of
CP asymmetries.

For $B^-\rightarrow D^-D^{0*}$, $D^{+*}D^0$ and $\bar{B}_d^0\rightarrow
\bar{K^0}\eta$, the electroweak penguin contributions to CP asymmetries are
dominant, they can enhance the CP asymmetries by an order of magnitude.

For $B^-\rightarrow K^-\eta$, $\rho^0\phi^-$, $K^-\omega$, $K^-\rho^0$,
$K^{-*}\eta$ and $\bar{B}_d^0\rightarrow \bar{K^0}\phi^0$, $\bar{K^0}\rho^0$,
$\bar{K^0}\omega$, $\bar{K^0}^*\pi^0$, the electroweak penguin contributions
are non-negligible or competable with $QCD$ and/or tree level contributions
in CP asymmetries, the percentage is larger than $30\%$.

For all decay modes, the CP asymmetries change not very much in magnitude
except from $-20.64\%$ to $-6.81$ for $B^-\rightarrow
K^-\rho^0$ and from $-35.07\%$ to $-4.85\%$ for $B^-\rightarrow \pi^-\rho^0$.

At last, we should mention that some decay modes have been calculated by
others using a similar approach, for example, $B^-\rightarrow \pi^-\phi$ in
ref.[13], $B^-\rightarrow \pi^- \overline {K^0}, K^-\phi$ in ref. [20,21].
Our results agree with them.

In summary, we can say that for some decay modes, the electroweak penguin
contributions are not negligible, they can enhance or reduce the QCD
and/or tree level contributions by  at least $30\%$, and can even play
dominant role in decay width, but can not  change the CP asymmetries
in magnitude largely.

\section*{Acknoledgement}

This work is supported in part by the National Natural Science Foundation
of China and the Grant of State Commission of Science and Technology of
China.

\section*{Appendix}
\vspace{0.3cm}
\subsection*{Parameters for the Numerical Calculations:}
In order to calculate branching ratios and CP-violating asymmetries, we
use the following values for the mass-parameters,
formfactors, decay constants, etc.
\vspace{0.3cm}

$\bullet$ quark masses:
$$ m_u=5MeV,~~~~m_d=10MeV,~~~m_c=1.35GeV,~~~~~~~~~~~~~~~$$
$$m_s=175MeV,~~m_b=5.0GeV,~~m_t=174GeV.~~~~~~~~~~~~~~~~$$

$\bullet$ meson masses and decay constants:

$M_{B_u^-}$~=~5.279GeV
\begin{center}
\begin{tabular}{|c|c|c|c|c|c|c|c|c|c|c|}\hline
meson (P) & $\pi^{\pm}$ & $\pi^0$ & $\eta$ & $\eta'$ & $K^{\pm}$ & $K^0$ &
$D^{\pm}$ & $D^0$ & $D_s^{\pm}$ & $\eta_c$\\
\hline
mass(MeV) & 140 & 135 & 547 & 958 & 494 & 498 & 1869 & 1865 & 1969 & 2979\\
\hline
decay(MeV) & & & $\eta_{u,d}:92$ & $\eta'_{u,d}:49$ & & & & & & \\
constant & 132 & 93 & $\eta_s:-105$ & $\eta'_s:120$ & 161 & 161 &
220 & 220 & 280 & 350\\
\hline
\end{tabular}
\end{center}
\vspace{0.13cm}
\begin{center}
[6~\begin{tabular}{|c|c|c|c|c|c|c|c|c|c|c|}
\hline
meson(V) & $\rho^{\pm}$ & $\rho^0$ & $\omega$ & $\phi$ & $K^{\pm^*}$ &
$K^{0^*}$ & $D^{\pm^*}$ & $D^{0^*}$ & $D_s^{\pm^*}$ & $J/\psi$\\
\hline
mass(MeV) & 770 & 770 & 782 & 1019 & 892 & 896 & 2010 & 2007 & 2110 & 3097\\
\hline
decay-const.(MeV)& 221 & 156 & 156 & 233 & 221 & 221 & 240 & 240 & 280 & 384\\
\hline
\end{tabular}
\end{center}

$\bullet$Pole masses (GeV):
\begin{center}
\begin{tabular}{|c|c|c|c|c|}
\hline
current & $m(0^-)$ & $m(1^-)$ & $m(0^+)$ & $m(1^+)$\\
\hline
${\bar d}c$ & 1.87 & 2.01 & 2.47 & 2.42\\
\hline
${\bar s}c$ & 1.97 & 2.11 & 2.60 & 2.53\\
\hline
${\bar u}b$ & 5.27 & 5.32 & 5.78 & 5.71\\
\hline
${\bar s}b$ & 5.38 & 5.43 & 5.89 & 5.82\\
\hline
${\bar c}b$ & 6.30 & 6.34 & 6.80 & 6.73\\
\hline
\end{tabular}
\end{center}
\vspace{0.3cm}

$\bullet$Form factors at zero momentum transfer:
\begin{center}
\begin{tabular}{|c|c|c|c|c|c|}
\hline
Decay & $F_1=F_0$ &$~~~V~~~$ &$~~A_1~~$ &$~~A_2~~$ &$A_3=A_0$\\
\hline
$B\rightarrow D$ & 0.690 & & & & \\
\hline
$B\rightarrow K$ & 0.379 & & & & \\
\hline
$B\rightarrow \pi$ & 0.333 & & & & \\
\hline
$B\rightarrow \eta$ & 0.307 & & & & \\
\hline
$B\rightarrow \eta'$ & 0.254 & & & & \\
\hline
$B\rightarrow D^*$ & & 0.705 & 0.651 & 0.686 & 0.623\\
\hline
$B\rightarrow K^*$ & & 0.369 & 0.328 & 0.331 & 0.321\\
\hline
$B\rightarrow \rho$ & & 0.329 & 0.283 & 0.283 & 0.281\\
\hline
$B\rightarrow \omega$ & & 0.328 & 0.281 & 0.281 & 0.280\\
\hline
\end{tabular}
\end{center}

\vspace{2cm}

\def\tmtwo{$\times10^{-2}$}\def\tmthree{$\times10^{-3}$}
\def\tmfour{$\times10^{-4}$}\def\tmfive{$\times10^{-5}$}
\def\tmsix{$\times10^{-6}$} \def\tmseven{$\times10^{-7}$}
\def\tmeight{$\times10^{-8}$}\def\tmnine{$\times10^{-9}$}
\def\tmtwelve{$\times10^{-12}$}\def\tmfiveteen{$\times10^{-15}$}
\def\tmsixteen{$\times10^{-16}$}\def\tmseventeen{$\times10^{-17}$}
\def\tmeighteen{$\times10^{-18}$}\def\tmnineteen{$\times10^{-19}$}
\def\tmtwenty{$\times10^{-20}$}  \def\tmtwentyone{$\times10^{-21}$}
\def\tmtwentyfive{$\times10^{-25}$}\def\tmthreeteen{$\times10^{-13}$}
\def\tmten{$\times10^{-10}$}
\begin{center}
Table 1{~~~~}   Branching ratios and CP asymmetries with $QCD+QED$ and\\
with only QCD penguin contributions in $ B^-\rightarrow  PP $
(through $ b \rightarrow  d $ transition)
\begin{tabular}{|l|c|c|c|c|c|c|}
\hline
    &\multicolumn{2}{|c|} {QCD+QED} &\multicolumn {2}{|c|}{QCD} &
\multicolumn {2}{|c|}{corrections}\\
\cline{2-7} decay mode  & Br & ${\cal A}_{cp} [\%] $ &
Br & ${\cal A}_{cp} [\%] $ &$\delta \Gamma/\Gamma [\%]$&$\delta {\cal A}/
     {\cal A} [\%]$ \\
\hline
$B^-\rightarrow \pi^-\pi^0 $
             & 4.88\tmsix & 0.7 & 5.15\tmsix & 0.6 & $-$5.2 & 11.9\\
\hline
$B^-\rightarrow \pi^-\eta$
             & 5.82\tmsix & 32.3 & 5.61\tmsix & 33.9 &3.8 &$-$4.7\\
\hline
$B^-\rightarrow \pi^-\eta'$
             & 9.57\tmsix & 15.1 & 9.64\tmsix & 15.0 & $-$0.7 & 0.5\\
\hline
$B^-\rightarrow \pi^-\eta_c $
             & 2.93\tmsix & 0 & 3.89\tmsix & 0 & $-$24.6 & \\
\hline
$B^-\rightarrow  K^-K^0 $
             & 7.20\tmseven & 2.9 & 7.33\tmseven & 2.9 & $-$1.8 & 1.0\\
\hline
$B^-\rightarrow D^-D^0 $ \
             & 4.17\tmfour & $-$2.0 & 4.18\tmfour & $-$1.9 & $-$0.2 & 0.6\\
\hline
\end{tabular}
\end{center}

\begin{center}
Table 2{~~~~}   Branching ratios and CP asymmetries with $QCD+QED$ and\\
	 with only QCD penguin contributions in $ B^-\rightarrow  PP $
	 (through $ b \rightarrow  s $ transition)
\begin{tabular}{|l|c|c|c|c|c|c|}
\hline
    &\multicolumn{2}{|c|} {QCD+QED} &\multicolumn {2}{|c|}
{QCD} & \multicolumn{2}{|c|} {corrections} \\
\cline{2-7} decay mode & Br & ${\cal A}_{cp}[\%]$ &  Br & ${\cal A}_{cp}[\%]$&
$\delta \Gamma/\Gamma[\%]$ & $\delta {\cal A}/{\cal A}[\%]$\\
\hline
$B^-\rightarrow K^-\pi^0 $
            &7.21\tmsix & $-$6.3 &4.98\tmsix &$-$8.8 &44.7 & $-$28.5\\
\hline
$B^-\rightarrow K^-\eta$
            &2.57\tmseven & 29.6 &4.06\tmseven & 17.4 &$-$36.8 & 69.9\\
\hline
$B^-\rightarrow K^-\eta'$
            &1.30\tmfive &$-$3.3 &1.37\tmfive &$-$3.1 &$-$5.8 & 5.7\\
\hline
$B^-\rightarrow K^-\eta_c $ &7.34\tmfive &0&9.43\tmfive &0&$-$22.2 & \\
\hline
$B^-\rightarrow \overline{K^0}\pi^- $
            &8.25\tmsix &$-$0.2 &8.40\tmsix &$-$0.2&$-$1.8 & 0.9\\
\hline
$B^-\rightarrow D^0D_s^- $ &1.45\tmtwo &0.09 &1.45\tmtwo &0.09 &$-$0.2 & 0.6\\
\hline
\end{tabular}
\end{center}
\begin{center}
Table 3{~~~~}   Branching ratios and CP asymmetries with $QCD+QED$ and\\
   with only QCD penguin contributions in $ B^-\rightarrow  PV $
   (through $ b \rightarrow  d $ transition)
\begin{tabular}{|l|c|c|c|c|c|c|}
\hline
    &\multicolumn{2}{|c|} {QCD+QED} &\multicolumn {2}{|c|}{QCD} &
\multicolumn {2}{|c|}{corrections}\\
\cline{2-7} decay mode  & Br & ${\cal A}_{cp}[\%]$  & Br & ${\cal A}_{cp}[\%]$&
$\delta \Gamma/\Gamma[\%]$ & $\delta {\cal A}/{\cal A}[\%]$\\
\hline
$B^-\rightarrow \rho^-\eta$
          &1.12\tmfive &$-$5.5 &1.15\tmfive &$-$5.4 &$-$2.6 &2.2\\
\hline
$B^-\rightarrow \rho^-\eta'$
          &1.14\tmfive &$-$6.6 &1.12\tmfive &$-$6.7 &1.4 & $-$1.6 \\
\hline
$B^-\rightarrow \rho^-\pi^0 $
          &1.24\tmfive &3.2 & 1.28\tmfive &3.1 &$-$2.8 & 3.9\\
\hline
$B^-\rightarrow \rho^-\eta_c $
          &9.36\tmseven &0 &1.24\tmsix &0 &$-$24.6 & \\
\hline
$B^-\rightarrow \omega\pi^- $
          &3.49\tmsix &12.3&3.56\tmsix &11.9&$-$1.9 & 3.4\\
\hline
$B^-\rightarrow \rho^0\pi^- $
           &4.20\tmsix &$-$4.9&4.69\tmsix &$-$35.1&$-$10.4 & $-$86.2\\
\hline
$B^-\rightarrow \phi\pi^- $
            &6.17\tmnine &0.2&1.19\tmtwelve &0 &$5.2\times10^5$ & \\
\hline
$B^-\rightarrow \pi^-J/\psi $
            &1.29\tmsix &$-$0.3 &1.78\tmsix &0&$-$27.7 & \\
\hline
$B^-\rightarrow D^{-*}D^0 $
            &3.58\tmfour &$-$0.7&3.59\tmfour &$-$0.7&$-$0.3 & 0.7\\
\hline
$B^-\rightarrow D^-D^{0*} $
            &2.49\tmfour &$-$2.3&2.62\tmfour &$-$0.1&$-$4.7  &
$1.8\times10^3$\\
\hline
$B^-\rightarrow K^{-*}K^0 $
            &1.73\tmeight &5.8 &1.95\tmeight &5.3 &$-$11.1 & 8.5\\
\hline
$B^-\rightarrow K^-K^{0*} $
             &4.38\tmseven &3.4 &4.55\tmseven &3.3 &$-$3.6 & 2.2\\
\hline
\end{tabular}
\end{center}

\begin{center}
Table 4{~~~~}   Branching ratios and CP asymmetries with $QCD+QED$ and\\
   with  only QED penguin contributions in $ B^-\rightarrow  PV $
(through $ b \rightarrow  s $ transition)
\begin{tabular}{|l|c|c|c|c|c|c|}
\hline
    &\multicolumn{2}{|c|} {QCD+QED} &\multicolumn {2}{|c|}
{QCD} &\multicolumn {2}{|c|}{corrections}\\
\cline{2-7} decay mode  & Br & ${\cal A}_{cp}[\%]$ & Br & ${\cal A}_{cp}[\%]$&
$\delta \Gamma/\Gamma[\%]$ & $\delta {\cal A}/{\cal A}[\%]$\\
\hline
$B^-\rightarrow K^-\omega $
            &6.83\tmseven&$-$13.1 &3.70\tmseven&$-$20.9 &84.7 & $-$37.2\\
\hline
$B^-\rightarrow K^-\rho^0 $
         &1.56\tmsix&$-$6.8 &3.79\tmseven&$-$20.6 &$3.1\times10^2$ & $-$67.0\\
\hline
$B^-\rightarrow K^-\phi $
              &5.50\tmsix &$-$0.3&7.42\tmsix &$-$0.2&$-$25.8 & 19.5\\
\hline
$B^-\rightarrow K^-J/\psi $
              &3.19\tmfive &0 &4.24\tmfive &0 &$-$24.8 & \\
\hline
$B^-\rightarrow K^{-*}\eta $
             &1.06\tmsix &$-$11.2 &5.90\tmseven &$-$18.3 &79.1 & $-$38.5\\
\hline
$B^-\rightarrow K^{-*}\eta' $
             &6.68\tmeight &$-$58.5 &4.20\tmeight &$-$69.1&59.1 & $-$15.3\\
\hline
$B^-\rightarrow K^{-*}\pi^0 $
              &5.33\tmsix &$-$9.4 &3.67\tmsix &$-$13.0 &45.3 & $-$27.9\\
\hline
$B^-\rightarrow K^{-*}\eta_c $
               &2.26\tmfive &0 &2.91\tmfive &0 &$-$22.2 & \\
\hline
$B^-\rightarrow \overline{K^0}^*\pi^- $
               &5.18\tmsix &$-$0.2 &5.38\tmsix &$-$0.2 &$-$3.6 & 2.2\\
\hline
$B^-\rightarrow \overline{K^0} \rho^- $
               &1.73\tmseven &$-$0.4 &1.97\tmseven &$-$0.3 &$-$12.2 & 9.0 \\
\hline
$B^-\rightarrow D_s^{-*}D^0 $
               &9.50\tmthree&0.03 &9.53\tmthree&0.03 &$-$0.3 & 0.7\\
\hline
$B^-\rightarrow D_s^-D^{0*} $
               &8.12\tmthree &0.006 &8.15\tmthree &0.006&$-$0.3 & 0.7\\
\hline
\end{tabular}
\end{center}

\begin{center}
Table 5{~~~~}   Branching ratios and CP asymmetries with $QCD+QED$ and\\
with only QCD penguin contributions in $ \overline{B^0_d}\rightarrow  PP $
(through $ b \rightarrow  d $ transition)
\begin{tabular}{|l|c|c|c|c|c|c|}
\hline
    &\multicolumn{2}{|c|} {QCD+QED} &\multicolumn {2}{|c|}{QCD} &
\multicolumn {2}{|c|}{corrections}\\
\cline{2-7} decay mode  & Br & ${\cal A}_{cp} [\%] $ &
Br & ${\cal A}_{cp} [\%] $ &$\delta \Gamma/\Gamma [\%]$ &
$\delta {\cal A}/{\cal A}[\%]$\\
\hline
$ \overline{B^0_d}\rightarrow \pi^+  \pi^-$
        & 7.95\tmsix & $-$42.5 & 7.98\tmsix & $-$42.7 & $-$0.3 & $-$0.4\\
\hline
$ \overline{B^0_d}\rightarrow \pi^0  \pi^0 $
   & 6.09\tmeight & $-$18.9 &1.00\tmseven & $-$16.9 & $-$39.1 & 11.7  \\
\hline
$ \overline{B^0_d}\rightarrow \pi^0  \eta $
          & 2.83\tmsix & 0.9 & 2.84\tmsix & 0.9 & $-$0.3 & 0.2\\
\hline
$ \overline{B^0_d}\rightarrow \pi^0  \eta' $
          & 5.27\tmsix & 0.4 & 5.46\tmsix & 0.4 & $-$3.3 & $-$2.0\\
\hline
$ \overline{B^0_d}\rightarrow \pi^0  \eta_c $
           & 1.43\tmsix & 26.1 & 1.89\tmsix & 23.0 & $-$24.6 & 13.9 \\
\hline
$ \overline{B^0_d}\rightarrow  \eta  \eta $
           & 1.25\tmsix & 3.8 & 1.18\tmsix & 3.8 & 6.5 & $-$1.1\\
\hline
$ \overline{B^0_d}\rightarrow \eta  \eta' $
           & 5.11\tmsix & 2.1 & 5.06\tmsix & 2.1 & 1.1 & 0.03\\
\hline
$ \overline{B^0_d}\rightarrow \eta  \eta_c $
           & 7.70\tmseven & 26.1 &  1.02\tmsix & 23.0 & $-$24.6 & 13.9\\
\hline
$ \overline{B^0_d}\rightarrow \eta'  \eta' $
           & 8.45\tmseven & 1.4 & 8.54\tmseven & 1.4 & $-$1.0 & 0.2 \\
\hline
$ \overline{B^0_d}\rightarrow \eta'  \eta_c $
            & 2.35\tmseven & 26.1 & 3.11\tmseven & 23.0 & $-$24.6 & 13.9\\
\hline
$ \overline{B^0_d}\rightarrow K^0  \overline{K^0} $
           & 7.01\tmseven & 0.02 & 7.14\tmseven & 0.03 & $-$1.8 & $-$25.7 \\
\hline
$ \overline{B^0_d}\rightarrow  D^+  D^- $
               & 4.05\tmfour & 30.2 & 4.05\tmfour & 30.2 & $-$0.2 & 0.1 \\
\hline
\end{tabular}\end{center}
\begin{center}
Table 6{~~~~}   Branching ratios and CP asymmetries with $QCD+QED$ and\\
with only QCD penguin contributions in $ \overline{B^0_d}\rightarrow  PP $
(through $ b \rightarrow  s $ transition)
\begin{tabular}{|l|c|c|c|c|c|c|}
\hline
    &\multicolumn{2}{|c|} {QCD+QED} &\multicolumn {2}{|c|}{QCD} &
     \multicolumn {2}{|c|}{corrections}\\
\cline{2-7} decay mode  & Br & ${\cal A}_{cp} [\%] $ &
Br & ${\cal A}_{cp} [\%] $ &$\delta \Gamma/\Gamma [\%]$ &
$\delta {\cal A}/{\cal A} [\%]$ \\
\hline
$ \overline{B^0_d}\rightarrow  K^-\pi^+ $
     & 9.89\tmsix & $-$7.4 & 9.56\tmsix & $-$7.6 & 3.4 & $-$2.7\\
\hline
$ \overline{B^0_d}\rightarrow   \overline{K^0}\pi^0 $
     & 2.57\tmsix & 0.8 & 4.08\tmsix & 0.5 & $-$37.0 & 68.3  \\
\hline
$ \overline{B^0_d}\rightarrow  \overline{K^0} \eta  $
      & 6.38\tmeight & 7.4 & 2.89\tmseven & 1.5 & $-$77.9 & $3.9\times10^2$\\
\hline
$ \overline{B^0_d}\rightarrow  \overline{K^0}\eta'  $
      &  1.22\tmfive & $-$0.5 & 1.31\tmfive & $-$0.4 & $-$7.0 & 6.0\\
\hline
$ \overline{B^0_d}\rightarrow \overline{K^0} \eta_c  $
       &  7.15\tmfive & 0 & 9.18\tmfive & 0 & $-$22.2 &  \\
\hline
$ \overline{B^0_d}\rightarrow D^+ + D_s^- $
       &  1.41\tmtwo & 0 & 1.41\tmtwo & 0 & $-$0.2 &     \\
\hline
\end{tabular}\end{center}
\begin{center}
Table 7{~~~~}   Branching ratios and CP asymmetries with $QCD+QED$ and\\
with only QED penguin contributions in $ \overline{B^0_d}\rightarrow  PV $
(through $ b \rightarrow  d $ transition)
\begin{tabular}{|l|c|c|c|c|c|c|}
\hline
    &\multicolumn{2}{|c|} {QCD+QED} &\multicolumn {2}{|c|}{QCD} &
     \multicolumn{2}{|c|} {corrections}\\
\cline{2-7} decay mode  & Br & ${\cal A}_{cp} [\%] $ &
Br & ${\cal A}_{cp} [\%] $ &$\delta \Gamma/\Gamma [\%]$
&$\delta {\cal A}/{\cal A} [\%]$ \\
\hline
$ \overline{B^0_d}\rightarrow \pi^0 + \rho^0 $
    & 1.27\tmseven & $-$26.3 & 2.73\tmseven & $-$24.1& $-$53.6 & 9.2 \\
\hline
$ \overline{B^0_d}\rightarrow \pi^0  \omega $
     & 1.09\tmseven & 10.5 & 1.24\tmseven & 10.7 & $-$12.4 & $-$2.5 \\
\hline
$ \overline{B^0_d}\rightarrow \eta  \rho^0 $
     & 6.41\tmseven & 3.5 & 5.59\tmseven & 3.6 & 14.6 & $-$4.3 \\
\hline
$ \overline{B^0_d}\rightarrow \eta'  \rho^0 $
      & 2.93\tmsix &1.1 & 2.70\tmsix & 1.1 & 8.5 & $-$1.1  \\
\hline
$ \overline{B^0_d}\rightarrow \eta  \omega $
       &  6.74\tmseven & $-$14.5 & 7.66\tmseven & $-$13.9 & $-$12.0 &  4.8   \\
\hline
$ \overline{B^0_d}\rightarrow \eta'  \omega $
   & 2.92\tmsix & $-$3.7 & 2.89\tmsix & $-$3.7 & 1.0 & $-$0.3  \\
\hline
$ \overline{B^0_d}\rightarrow \eta_c  \rho^0 $
      &  4.56\tmseven & 26.1 & 6.05\tmseven & 23.0& $-$24.6 & 13.9       \\
\hline
$ \overline{B^0_d}\rightarrow \eta_c  \omega $
       &   4.50\tmseven & 26.1 & 5.96\tmseven & 23.0& $-$24.6 & 13.9   \\
\hline
$ \overline{B^0_d}\rightarrow \phi  \pi^0 $
     & 3.01\tmnine & 0 & 5.8\tmthreeteen &0 & $5.2\times10^5$ &  \\
\hline
$ \overline{B^0_d}\rightarrow \phi  \eta $
    & 1.65\tmnine & 0 & 3.17\tmthreeteen & 0 & $5.2\times10^5$ &    \\
\hline
$ \overline{B^0_d}\rightarrow \phi  \eta' $
    &  5.22\tmten & 0 & 1.00\tmthreeteen & 0 & $5.2\times10^5$&     \\
\hline
$ \overline{B^0_d}\rightarrow \rho^+  \pi^- $
       &  5.88\tmsix & 2.2 & 5.91\tmsix & 2.09 & $-$0.5 & 2.8 \\
\hline
$ \overline{B^0_d}\rightarrow \rho^-  \pi^+ $
         &  2.28\tmfive & 4.8 & 2.29\tmfive & 4.8 & $-$0.4 & 0.9 \\
\hline
$ \overline{B^0_d}\rightarrow \pi^0  J/\psi $
     & 6.29\tmseven & 30.4 & 8.69\tmseven &26.2 & $-$27.7 & 16.0   \\
\hline
$ \overline{B^0_d}\rightarrow \eta  J/\psi $
    &  3.23\tmseven & 30.4 & 4.47\tmseven &26.2 & $-$27.7 & 16.0    \\
\hline
$ \overline{B^0_d}\rightarrow \eta'  J/\psi $
    &  8.72\tmseven & 30.4 & 12.06\tmseven & 26.2 & $-$27.7 & 16.0      \\
\hline
$ \overline{B^0_d}\rightarrow K^0  \overline{K^0}^* $
       & 1.68\tmeight & 5.8 & 1.89\tmeight & 5.3 & $-$11.1 & 8.5  \\
\hline
$ \overline{B^0_d}\rightarrow K^{0*}  \overline{K^0} $
            &  4.27\tmseven & 3.4 & 4.42\tmseven & 3.3 & $-$3.6 & 2.2  \\
\hline
$ \overline{B^0_d}\rightarrow D^{+*}  D^- $
   & 2.42\tmfour & $-$2.3 & 2.54\tmfour & $-$0.1 & $-$4.7 & $1.8\times10^3$  \\
\hline
$ \overline{B^0_d}\rightarrow D^+  D^{-*} $
      &  3.47\tmfour & $-$0.7 & 3.48\tmfour & $-$0.7 & $-$0.3 & 0.7  \\
\hline
\end{tabular}\end{center}
\begin{center}
Table 8{~~~~}   Branching ratios and CP asymmetries with $QCD+QED$ and\\
with only QED penguin contributions in $ \overline{B^0_d}\rightarrow  PV $
(through $ b \rightarrow  s $ transition)
\begin{tabular}{|l|c|c|c|c|c|c|}
\hline
    &\multicolumn{2}{|c|} {QCD+QED} &\multicolumn {2}{|c|}{QCD} &
\multicolumn{2}{|c|} {corrections}\\
\cline{2-7} decay mode  & Br & ${\cal A}_{cp} [\%] $ &
Br & ${\cal A}_{cp} [\%] $ &$\delta \Gamma/\Gamma [\%]$ &
$\delta {\cal A}/{\cal A} [\%]$ \\
\hline
$ \overline{B^0_d}\rightarrow K^- \rho^+  $
     &  6.91\tmseven & $-$19.7 & 6.16\tmseven & $-$21.5 & 12.1 & $-$8.3 \\
\hline
$ \overline{B^0_d}\rightarrow K^{-*} \pi^+  $
     & 7.48\tmsix & $-$12.2 & 7.03\tmsix & $-$12.8 & 6.4 & $-$4.9 \\
\hline
$ \overline{B^0_d}\rightarrow \overline{K^0}^*  \pi^0 $
     & 1.59\tmsix & 0.9 & 2.61\tmsix & 0.5 & $-$39.0 & 74.3 \\
\hline
$ \overline{B^0_d}\rightarrow  \overline{K^0}  \rho^0 $
     &  1.93\tmseven & 5.1 & 9.89\tmeight & 11.9 & 95.4 & $-$57.3 \\
\hline
$ \overline{B^0_d}\rightarrow  \overline{K^0}  \omega $
    &2.62\tmseven & $-$4.3 &9.54\tmeight & $-$10.5 & $1.7\times10^2$ &
$-$58.7\\
\hline
$ \overline{B^0_d}\rightarrow  \overline{K^0}^*  \eta $
            &  4.03\tmsix & 0.3 & 4.96\tmsix & 0.2 & $-$18.8 & 28.4\\
\hline
$ \overline{B^0_d}\rightarrow  \overline{K^0}^*  \eta' $
             & 3.29\tmsix & 0.05 & 2.84\tmsix &0.06 & 15.7 &$-$18.4 \\
\hline
$ \overline{B^0_d}\rightarrow  \overline{K^0}^*  \eta_c $
              &  2.20\tmfive & 0 & 2.83\tmfive & 0& $-$22.2 &\\
\hline
$ \overline{B^0_d}\rightarrow  \overline{K^0}  \phi $
              & 5.36\tmsix & $-$0.3 & 7.22\tmsix & $-$0.2 & $-$25.8 & 19.5  \\
\hline
$ \overline{B^0_d}\rightarrow  \overline{K^0}  J/\psi $
              & 3.10\tmfive & 0.01 & 4.13\tmfive & 0 & $-$24.8 &      \\
\hline
$ \overline{B^0_d}\rightarrow  D^{+*}  D_s^- $
              & 7.89\tmthree & 0.006 &7.91\tmthree & 0.006 & $-$0.3 & 0.7 \\
\hline
$ \overline{B^0_d}\rightarrow  D^+  D_s^{-*} $
              & 9.22\tmthree & 0.03 & 9.24\tmthree & 0.03 & $-$0.3 & 0.7   \\
\hline
\end{tabular}\end{center}

\end{document}